\documentclass[a4paper]{jpconf}
\usepackage{graphicx}
\usepackage{amssymb}
\usepackage{cite}
\begin{document}
\title{Unflavored Leptogenesis and Neutrino Masses in Flavored SUSY SU(5) model}

\author{M.A. Loualidi\footnote{Speaker} and M. Miskaoui}

\address{LPHE, Modeling and Simulations, Faculty of Science,
	Mohammed V University in Rabat, 10090 Rabat, Morocco}

\ead{mr.medamin@gmail.com}

\begin{abstract}
We propose a model with $D_4$ flavor symmetry for leptogenesis in the framework of supersymmetric $SU(5)$ grand unified theory. 
Neutrino masses arise from the type I seesaw mechanism where the spontaneous symmetry breaking of the flavor symmetry leads to the well-known trimaximal mixing ($\rm{TM}_2$). We find that this model predicts the normal hierarchy (NH) for neutrino masses. We also find that the predicted range of the effective neutrino mass $%
m_{\beta \beta }$ can be tested at future neutrinoless double beta decay
experiments. For the generation of the baryon
asymmetry of the Universe (BAU), we consider the unflavored leptogenesis regime and we take into account the decay of
the three righ-handed neutrinos $N_{i}$. We conclude that the BAU
originates from a correction to the Dirac operator giving rise to a new coupling and a $CP$ phase which both play an important role in generating the observed BAU.
\end{abstract}

\section{Introduction}
The discovery of neutrino oscillations is the first direct experimental evidence for new physics beyond
the standard model (SM) \cite{R1}. One of the
simplest ways to probe the origin of the small neutrino
masses is by invoking the type I seesaw mechanism in which heavy
right-handed neutrinos (RHNs) are added to the SM \cite{R2}. An attractive aspect of this mechanism is that it adresses three major puzzles in particle physics: the nature of
neutrinos, the smalness of neutrino masses, and the matter-antimatter
asymmetry problem. These three aspects provide also a
motivation for leptogenesis as an approach to explain the
BAU \cite{R3}. In a nutshell, this approach requires lepton number violation which emerges naturally in type I seesaw models. Then, a
lepton asymmetry is created by the out of-thermal-equilibrium and $CP$
violating decays of the RHNs that is converted into a baryon
asymmetry by electroweak sphalerons \cite{R4}.\newline
In this context, many models have been proposed in recent years, and since
the seesaw scale is associated with the
masses of RHNs which are generally assumed to be around the
unification scale, supersymmetric grand unified theories (SUSY-GUTs) are one
of the most appealing models to study neutrinos and their properties \cite{R5}. On the other hand, neutrino mixing can be
described efficiently by using some non-Abelian discrete symmetry such as
the $A_{4}$ group which is broadly used in $SU(5)$ models to study the
patterns of neutrino mixing; see for example Refs. \cite%
{R6}.

In this letter, based on our recent work \cite{R21}, we will discuss the
generation of the BAU via the leptogenesis mechanism in the framework of a
SUSY $SU(5)$ model extended by a $D_{4}\times U(1)$ symmetry. The dihedral
group $D_{4}$ is introduced to adress the neutrino flavor structure while $%
U(1)$ controls the invariance of the superpotential of the model. We present
here an update of the study done in Ref. \cite{R21} where we assume that the
RH neutrino mass spectrum is not strongly hierarchical which requires taking
into account the contribution of the decay of the three RHNs. In
this case, we find that the $CP$ asymmetries for each Majorana neutrino $%
\varepsilon _{N_{i}}$ is dominated by the contribution of a next-to-leading
(NLO) order correction to the usual type I seesaw terms. In particular, we
find that the correlation between the baryon asymmetry parameter $Y_{B}$
and the parameters from the NLO term satisfies the experimental bound of $%
Y_{B}$ from the Planck collaboration \cite{R22}. We find also that the
predicted values of the effective Majorana mass $m_{\beta \beta }$\ measured
in neutrinoless double beta decay ($0\nu \beta \beta $) are testable at
future decay experiments.

\section{Neutrino sector in $D_{4}\times SU(5)$ model}

The neutrino sector of the present model is enlarged by adding three gauge-singlet RHNs $N_{i=1,2,3}^{c}$ charged under $D_{4}\times U(1)$ group. These RHNs are responsible for generating the small neutrino masses via the type I seesaw mechanism as well as the BAU through the leptogenesis mechanism. The charges under $D_{4}\times U(1)$ of matter, Higgs and RHNs are listed in table (\ref%
{t1}). In this letter, the focus is put on the neutrino sector and
especially the study of the BAU; however, we should mention that in Ref.
\cite{R21} where the charged fermion sector is studied as well, the Higgs
multiplets $H_{\overline{45}}$ and $H_{24}$ were also used to produce the
following ratios of the Yukawa couplings of the first and second generations
$y_{e}/y_{d}=4/9$ and $y_{\mu }/y_{s}=9/2$ which are in perfect agreement
with experimental data \cite{R23}.
\begin{table}[h]
	\centering
	\begin{tabular}{l||l|l|l|l|l|l|l|l|l|l|l}
		\hline
		& $T_{1}$ & $T_{2}$ & $T_{3}$ & $F_{1}$ & $F_{2,3}$ & $N_{1}^{c}$ & $%
		N_{3,2}^{c}$ & $H_{5}$ & $H_{\overline{5}}$ & $H_{\overline{45}}$ & $H_{24}$
		\\ \hline
		$SU(5)$ & $10_{1}$ & $10_{2}$ & $10_{3}$ & $\overline{5}_{1}$ & $\overline{5}
		_{2,3}$ & $1_{1}^{\nu }$ & $1_{3,2}^{\nu }$ & $5_{H_{u}}$ & $\overline{5}
		_{H_{d}}$ & $\overline{45}_{H}$ & $24_{H}$ \\ \hline
		$D_{4}$ & $1_{+,-}$ & $1_{+,-}$ & $1_{+,+}$ & $1_{+,+}$ & $2_{0,0}$ & $%
		1_{+,+}$ & $2_{0,0}$ & $1_{+,+}$ & $1_{+,-}$ & $1_{+,+}$ & $1_{+,+}$ \\
		\hline
		$U\mathbb{(}1\mathbb{)}$ & $6$ & $12$ & $4$ & $13$ & $13$ & $-5$ & $-5$ & $%
		-8 $ & $4$ & $-16$ & $0$ \\ \hline
	\end{tabular}%
	\caption{The $SU(5)\times D_{4}$ representations and $U(1)$ charges of the
		matter, RH neutrinos and Higgs superfields. See Ref. \cite{R21} for the notation of the different fields.}
	\label{t1}
\end{table}
Furthermore, the scalar sector is extended by five flavons
needed for flavor symmetry breaking, $U(1)$ invariance and to structure
the neutrino mass matrix, see table (\ref{t2}) for their $%
D_{4}\times U(1)$ charge assignments.
\begin{table}[h]
	\centering
	\begin{tabular}{l||l|l|l|l|l}
		\hline
		Flavons & $\rho _{1}$ & $\rho _{2}$ & $\rho _{3}$ & $\digamma $ & $\Gamma $
		\\ \hline
		$D_{4}$ & $1_{+,+}$ & $1_{+,-}$ & $1_{-,-}$ & $2_{0,0}$ & $2_{0,0}$ \\ \hline
		$U\mathbb{(}1\mathbb{)}$ & $10$ & $10$ & $10$ & $10$ & $10$ \\ \hline
	\end{tabular}%
	\caption{The $D_{4}\times U(1)$ quantum numbers of the flavons used in the
		neutrino sector.}
	\label{t2}
\end{table}
Thus, by using the charge assignments in tables (\ref{t1}) and (\ref{t2}),
the superpotential invariant under the $SU(5)\times D_{4}\times U(1)$ group
is given by%
\begin{eqnarray}
	\mathcal{W}_{\nu }
	&=&y_{1}N_{1}^{c}F_{1}H_{5}+y_{2}N_{3,2}^{c}F_{2,3}H_{5}+y_{3}N_{1}^{c}N_{1}^{c}\rho _{1}+y_{4}N_{3,2}^{c}N_{3,2}^{c}\rho _{1}+y_{5}N_{1}^{c}N_{3,2}^{c}\digamma
	\nonumber \\
	&&+y_{6}N_{1}^{c}N_{3,2}^{c}\Gamma +y_{7}N_{3,2}^{c}N_{3,2}^{c}\rho
	_{2}+y_{8}N_{3,2}^{c}N_{3,2}^{c}\rho _{3}  \label{wn}
\end{eqnarray}%
where $y_{i=1,...,8}$ are Yukawa coupling constants. The first two terms give rise to the Dirac mass matrix $%
m_{D}$ while the rest of the terms induce the Majorana mass matrix $m_{M}$.
In particular, the couplings involving $\rho _{1}$ and $\digamma $ lead to
the well-known tribimaximal mixing (TBM) matrix \cite{R24}, while the couplings
incorporating $\rho _{2}$, $\rho _{3}$ and $\Gamma $ lead to $\rm{TM}_2$ which is consistent with the data. We assume that the VEVs of the flavon fields point in the following directions%
\begin{equation}
	\left\langle \rho _{1}\right\rangle =\upsilon _{\rho _{1}}\quad ,\quad
	\left\langle \rho _{2}\right\rangle =\upsilon _{\rho _{2}}\quad ,\quad
	\left\langle \rho _{3}\right\rangle =\upsilon _{\rho _{3}}\quad ,\quad
	\left\langle \digamma \right\rangle =(\upsilon _{\digamma },\upsilon
	_{\digamma })^{T}\quad ,\quad \left\langle \Gamma \right\rangle =(0,\upsilon
	_{\Gamma })^{T}  \label{va}
\end{equation}%
while the Higgs doublet develops its VEV as is customary $\left\langle
H_{u}\right\rangle =\upsilon _{u}$. Then, using the tensor product of $D_{4}$
irreducible representations (see eqs. (C.2) and (C.3) of Ref. \cite%
{R21}), we find that the Dirac and Majorana mass matrices have the following
forms%
\begin{equation}
	m_{D}=\upsilon _{u}Y_{D}=\upsilon _{u}\left(
	\begin{array}{ccc}
		y_{1} & 0 & 0 \\
		0 & y_{2} & 0 \\
		0 & 0 & y_{2}%
	\end{array}%
	\right) \quad ,\quad m_{M}=\left(
	\begin{array}{ccc}
		y_{3}\upsilon _{\rho _{1}} & y_{5}\upsilon _{\digamma } & y_{5}\upsilon
		_{\digamma }+y_{6}\upsilon _{\Gamma } \\
		y_{5}\upsilon _{\digamma } & y_{7}\upsilon _{\rho _{2}}-y_{8}\upsilon _{\rho
			_{3}} & 2y_{4}\upsilon _{\rho _{1}} \\
		y_{5}\upsilon _{\digamma }+y_{6}\upsilon _{\Gamma } & 2\lambda _{4}\upsilon
		_{\rho _{1}} & y_{7}\upsilon _{\rho _{2}}+y_{8}\upsilon _{\rho _{3}}%
	\end{array}%
	\right)   \label{md}
\end{equation}%
The conditions for TBM mixing \cite{R25,R26,R27}, and its deviation to $\rm{TM}_2$ require the imposition of the following assumptions\footnote{%
	The validity of these assumptions is discussed in appendix D of Ref. \cite%
	{R21}.} on $m_{D}$ and $m_{M}$: $y_{1}=y_{2}$, $y_{3}\upsilon _{\rho _{1}}+y_{5}\upsilon
	_{\digamma }=2y_{4}\upsilon _{\rho _{1}}$, and $y_{8}\upsilon _{\rho
		_{3}}=-y_{7}\upsilon _{\rho _{2}}=y_{6}\upsilon _{\Gamma }/2  \label{as1}$.
Now, we can apply the usual seesaw formula $m_{\nu }=m_{D}m_{M}^{-1}m_{D}^{T}
$ which gives rise to the total neutrino mass matrix
\begin{equation}
	m_{\nu }=\frac{m_{0}}{P}\left(
	\begin{array}{ccc}
		-\left( a+b\right) ^{2} & \left( a+b\right) \left( b+\mathrm{k}\right)  &
		b^{2}-\mathrm{k}^{2}-b\left( \mathrm{k}-a\right)  \\
		\left( a+b\right) \left( b+\mathrm{k}\right)  & -\left( b+\mathrm{k}\right)
		^{2} & -a^{2}-ab+b^{2}+\mathrm{k}b \\
		b^{2}-\mathrm{k}^{2}-b\left( \mathrm{k}-a\right)  & -a^{2}-ab+b^{2}+\mathrm{k%
		}b & a\mathrm{k}-b^{2}%
	\end{array}%
	\right)   \label{mn}
\end{equation}%
where we have adopted the following notations to simplify the
parametrization of $m_{\nu }$%
\begin{eqnarray}
	m_{0} &=&\frac{\left( y_{1}\upsilon _{u}\right) ^{2}}{M_{R}}~~,~~P=\left(
	a+2b+\mathrm{k}\right) \left( a\mathrm{k}-a^{2}+b^{2}-\mathrm{k}^{2}\right)
	\nonumber \\
	~a &=&\frac{y_{3}\upsilon _{\rho _{1}}}{M_{R}}~~,~~b=\frac{y_{5}\upsilon
		_{\digamma }}{M_{R}}~~,~~c=\frac{2y_{4}\upsilon _{\rho _{1}}}{M_{R}}~~,~~%
	\mathrm{k}=\frac{y_{6}\upsilon _{\Gamma }}{M_{R}}
\end{eqnarray}%
with $M_{R}$ being the mass scale of the RH neutrinos. In order to satisfy $%
CP$ violation in the lepton sector, we can take without loss of
generality, only the parameter $k$ to be complex -- $k\rightarrow \left\vert
\mathrm{k}\right\vert e^{i\phi _{k}}$ where $\phi _{k}$\ is a $CP$\
violating phase. After breaking
the flavor symmetry, we find that the matrix
$m_{\nu }$ enjoys a remnant $Z_{2}$ symmetry called a magic symmetry
refering to the equality of the sum of each row and the sum of each column
of $m_{\nu }$ \cite{R28}. This property implies that $m_{\nu }$ is
diagonalized by the trimaximal mixing matrix $\mathcal{U%
}_{TM_{2}}$\ so that $m_{\nu }^{\mathrm{diag}}=\mathcal{U}_{TM_{2}}^{\dagger
}m_{\nu }\mathcal{U}_{TM_{2}}$\ with%
\begin{equation}
	\mathcal{U}_{TM_{2}}=\left(
	\begin{array}{ccc}
		\sqrt{\frac{2}{3}}\cos \theta  & \frac{1}{\sqrt{3}} & \sqrt{\frac{2}{3}}\sin
		\theta e^{-i\sigma } \\
		-\frac{\cos \theta }{\sqrt{6}}-\frac{\sin \theta }{\sqrt{2}}e^{i\sigma } &
		\frac{1}{\sqrt{3}} & \frac{\cos \theta }{\sqrt{2}}-\frac{\sin \theta }{\sqrt{%
				6}}e^{-i\sigma } \\
		-\frac{\cos \theta }{\sqrt{6}}+\frac{\sin \theta }{\sqrt{2}}e^{i\sigma } &
		\frac{1}{\sqrt{3}} & -\frac{\cos \theta }{\sqrt{2}}-\frac{\sin \theta }{%
			\sqrt{6}}e^{-i\sigma }%
	\end{array}%
	\right)   \label{3-11}
\end{equation}%
where $\theta $ and $\sigma $ are respectively an arbitrary angle and a
phase. The complete mixing matrix for the neutrino sector is given by $\mathcal{U}_{\nu
}=\mathcal{U}_{TM_{2}}\mathcal{U}_{P}$ where $\mathcal{U}_{P}=\mathrm{diag}%
(1,e^{i\frac{\alpha _{21}}{2}},e^{i\frac{\alpha _{31}}{2}})$ is a diagonal
matrix that contains the two additional Majorana phases $\alpha _{21}$ and $%
\alpha _{31}$. The diagonalization of the neutrino matrix (\ref{mn}) by $%
\mathcal{U}_{TM_{2}}$ leads to the following eigenmasses
\begin{equation}
	\begin{array}{c}
		\left\vert m_{1}\right\vert =\frac{m_{0}}{\sqrt{(a-b)^{2}-\left\vert \mathrm{%
					\ k}\right\vert (a-b)\cos \phi _{k}+(\left\vert \mathrm{k}\right\vert ^{2}/4)%
		}}\quad ,\quad \left\vert m_{2}\right\vert =\frac{m_{0}}{\sqrt{%
				(a+2b)^{2}+2\left\vert \mathrm{k}\right\vert (a+2b)\cos \phi _{k}+\left\vert
				\mathrm{k}\right\vert ^{2}}} \\
		\left\vert m_{3}\right\vert =\frac{m_{0}}{\sqrt{(a+b)^{2}-\left\vert \mathrm{%
					\ k}\right\vert (a+b)\cos \phi _{k}+(\left\vert \mathrm{k}\right\vert ^{2}/4)%
		}}%
	\end{array}
	\label{mas}
\end{equation}%
where the denominators of $\left\vert m_{1}\right\vert $, $\left\vert
m_{2}\right\vert $ and $\left\vert m_{3}\right\vert $ corresponds to ratios
of the RHN masses $\left\vert M_{1}\right\vert $, $\left\vert
M_{2}\right\vert $ and $\left\vert M_{3}\right\vert $ and their mass scale $%
M_{R}$, respectively .Regarding the mixing angles in the case of trimaximal
mixing, they are expressed as as a function trimaximal mixing parameters $%
\sigma $ and $\theta $ as%
\begin{equation}
	\sin ^{2}\theta _{13}=\frac{2}{3}\sin ^{2}\theta ~,~\sin ^{2}\theta _{12}=%
	\frac{1}{3-2\sin ^{2}\theta }~,~\sin ^{2}\theta _{23}=\frac{1}{2}-\frac{%
		3\sin 2\theta }{2\sqrt{3}(3-2\sin ^{2}\theta )}\cos \sigma.
	\label{mix}
\end{equation}

\section{Leptogenesis}
The key component to apply the leptogenesis mechanism is
the presence of the RHNs $N_{i}^{c}$ as a means for the generation
of the small neutrino masses. The matter-antimatter asymmetry is
parametrized in terms of the baryon asymmetry parameter $Y_{B}$ which has been precisely measured by the
Planck collaboration: $Y_{B}=(8.72\pm 0.08)\times 10^{-11}$ \cite{R22}.

In order to dynamically generate this asymmetry, the three Sakharov
conditions must be fulfilled \cite{R29}: baryon number violation, $C$ and $CP$ violation and
departure from thermal equilibrium. In a type I seesaw scenario, a lepton
asymmetry $Y_{L}$ is generated
through the out-of-equilibrium $CP$ violating decays of $N_{i}^{c}$ that is
partially converted into the baryon asymmetry $Y_{B}$ via
sphaleron processes \cite{R4}. In order to estimate the value of $Y_{B}$, we
perform our study in the case of unflavored leptogenesis due to the fact
that all RHN masses are above $T=10^{12}(1+\tan ^{2}\beta )$ as
discussed in \cite{R21}. Moreover, since the RHN mass spectrum is not strongly hierarchical, we
consider the contributions of all three RHNs to the generation of
the BAU; $Y_{B}=\sum\limits_{i=1}^{3}Y_{Bi}$ with
\begin{equation}
	Y_{Bi}\approx -1.266\times 10^{-3}\varepsilon _{N_{i}}\eta _{ii}  \label{ybi}
\end{equation}%
where $\varepsilon _{N_{i}}$ is the $CP$ asymmetry parameter, $\eta _{ii}$ is the efficiency factor, while the numerical value in (%
\ref{ybi}) depends on the number densities of RHNs over the entropy
density and on the sphaleron transitions, see \cite{R21} for the details
yielding to this value.
The leading order terms given in eq. (\ref{wn}) induce higly suppressed $CP$
asymmetry parameter $\varepsilon _{N_{i}}$ and thus a suppression of the
value of $Y_{B}$. Thus, we have to rely on NLO terms to obtain an
appropriate suppression of the $CP$ asymmetries. Therefore, we consider a case
in which the NLO contribution arises in the neutrino sector and
only corrects $Y_{D}$ in eq. (\ref{md}). This is realized by
introducing a new flavon $\omega $ which transforms as $1_{+-}$ under $%
D_{4}$ with zero $U(1)$ charge
\begin{equation}
	\delta W_{D}=\frac{y_{9}}{\Lambda }N_{3,2}^{c}F_{2,3}H_{5}\omega   \label{D}
\end{equation}%
where $y_{9}$ is a complex coupling constant; $y_{9}=\left\vert
y_{9}\right\vert e^{i\phi _{\omega }}$. When the flavon field $\omega $
acquires its VEV as $\left\langle \omega \right\rangle =\upsilon _{\omega }$%
, we find the new Yukawa mass matrix\footnote{%
	Notice that the total light neutrino mass matrix involving the small
	correction $\delta Y_{D}$ is almost similar to the one in eq. (\ref{wn}) and
	yields approximately to the same neutrino phenomenology.}
\begin{equation}
	\mathcal{Y}_{D}=Y_{D}+\delta Y_{D}=\frac{m_{D}}{\upsilon _{u}}+\delta
	Y_{D}=\left(
	\begin{array}{ccc}
		y_{1} & 0 & 0 \\
		0 & y_{1} & 0 \\
		0 & 0 & y_{1}%
	\end{array}%
	\right) +\kappa e^{i\phi _{\omega }}\left(
	\begin{array}{ccc}
		0 & 0 & 0 \\
		0 & 0 & 1 \\
		0 & 1 & 0%
	\end{array}%
	\right)
\end{equation}%
where $\kappa =\frac{\left\vert y_{9}\right\vert \upsilon _{\omega }}{%
	\Lambda }$ is a free parameter which must be very small ---$\kappa <<1$---
to produce the correct BAU. The total Yukawa neutrino mass matrix is now
defined as $\mathcal{Y}_{\nu }=\mathcal{U}_{\nu }^{\dagger }\mathcal{Y}_{D}$%
. After we compute the product $\mathcal{Y}_{\nu }\mathcal{Y}_{\nu
}^{\dagger }$, we obtain the analytic expressions of the $CP$ asymmetries

\begin{eqnarray}
\varepsilon _{N_{1}} &\simeq &\frac{\kappa ^{2}\cos ^{2}\phi _{\omega }}{%
	9\pi }\left[ \cos ^{2}\left( \theta \right) \sin ^{2}\left( \frac{\alpha
	_{21}+4\phi _{\omega }}{2}\right) f\left( \frac{\tilde{m}_{2}}{\tilde{m}_{1}}%
\right) +2\sin ^{2}\left( \frac{\alpha _{21}-2\sigma }{2}\right) \sin
^{2}(2\theta )f\left( \frac{\tilde{m}_{3}}{\tilde{m}_{1}}\right) \right]
\nonumber \\
\varepsilon _{N_{2}} &\simeq &\frac{\kappa ^{2}\cos ^{2}\phi _{\omega }}{%
	9\pi }\left[ \cos ^{2}\left( \theta \right) \sin ^{2}\left( \frac{\alpha
	_{21}}{2}\right) f\left( \frac{\tilde{m}_{1}}{\tilde{m}_{2}}\right) +\sin
^{2}\left( \frac{\alpha _{21}-\alpha _{31}+2\sigma }{2}\right) \sin
^{2}\left( \theta \right) f\left( \frac{\tilde{m}_{3}}{\tilde{m}_{2}}\right) %
\right]  \\
\varepsilon _{N_{3}} &\simeq &\frac{\kappa ^{2}\cos ^{2}\phi _{\omega }}{%
	9\pi }\left[ 2\sin ^{2}(2\theta )\sin ^{2}\left( \frac{\alpha _{31}-2\sigma
}{2}\right) f\left( \frac{\tilde{m}_{1}}{\tilde{m}_{3}}\right) +\sin
^{2}\left( \frac{\alpha _{21}-\alpha _{31}+2\sigma }{2}\right) \sin
^{2}\left( \theta \right) f\left( \frac{\tilde{m}_{2}}{\tilde{m}_{3}}\right) %
\right]   \nonumber
\end{eqnarray}%
where $\tilde{m}_{i}$ are the washout mass parameters expressed as $\tilde{m}%
_{i}=\upsilon _{u}^{2}\frac{\left( \mathcal{Y}_{\nu }\mathcal{Y}_{\nu
	}^{\dagger }\right) _{ii}}{M_{i}}$. Notice here that the smallness of the
parameter $\kappa <<y_{1}$ implies that $\tilde{m}_{i}\approx m_{i}$. From the obtained expression of $%
\varepsilon _{N_{i}}$, we deduce that $Y_{B}$ in our model depends on the
trimaximal parameters, the active neutrino masses, the Majorana phases, as well as $%
\kappa $ and the phase $\phi _{\omega }$ coming from the NLO contribution to $Y_D$. The last parameter to evaluate in order to compute $Y_{Bi}
$ in eq. (\ref{ybi}) is the efficiency factor $\eta _{ii}$. We consider the
region of RHN masses smaller than $10^{14}$ $\mathrm{GeV}$,
preventing possible washout effects. In this case, $\eta _{ii}$ can be expressed as a function of the washout mass
parameter $\tilde{m}_{i}$ as \cite{R30}
\begin{equation}
	\eta _{ii}\approx \left( \frac{3.3\times 10^{-3}\textrm{eV}}{\tilde{m}%
		_{i}}+\left( \frac{\tilde{m}_{i}}{0.55\times 10^{-3}\textrm{eV}}%
	\right) ^{1.16}\right) ^{-1}\quad
\end{equation}%

\section{Numerical analysis and results}

From the first relation in eq. (\ref{mix}) and the $3\sigma $
range of $\theta _{13}$ from Ref. \cite{R31}, we deduce that $\theta $ lies in the range $0.1763\leq \theta
\leq 0.1920$. Moreover, by using the $3\sigma $ ranges of $\Delta m_{ij}^{2}=m_{i}^{2}-m_{j}^{2}$ in the case of the IH and the neutrino masses in eq. (\ref{mas}) as well
as the constraint on the sum of $m_i$ from cosmological
observations $\sum m_{i}<0.12$ $\mathrm{eV}$ \cite{R22}, we find that $%
\theta $ lies in the range $0.398\lesssim \theta \lesssim 0.579$ which
indicates that $\sin ^{2}\theta _{13}$ and $\sin ^{2}\theta _{12}$ fall far
outside their allowed $3\sigma $ experimental range. For this reason, the IH
pattern is excluded in our model.

The scale of neutrino masses can be probed by studying
the $0\nu \beta \beta $. The effective Majorana mass $\left\vert m_{\beta \beta }\right\vert
$ associated with this process is defined as $\left\vert m_{\beta \beta }\right\vert =\left\vert
\sum_{i}U_{ei}^{2}m_{i}\right\vert $ where $m_{i}$ are the neutrino
masses and $U_{ei}$ are the elements of the first row of the mixing matrix.
In this numerical analysis, we consider the current limits on $%
\left\vert m_{\beta \beta }\right\vert $ from KamLAND-Zen, CUORE, GERDA
and EXO experiments corresponding to $\left\vert m_{\beta \beta }\right\vert
<(0.061-0.165)\mathrm{eV}$, $\left\vert m_{\beta \beta
}\right\vert <(0.075-0.35)\mathrm{eV}$, $\left\vert m_{\beta
	\beta }\right\vert <(0.079-0.180)\mathrm{eV}$ and $\left\vert
m_{\beta \beta }\right\vert <(0.078-0.239)\mathrm{eV}$,
respectively \cite{R32}.
\begin{figure}[h]
	\centering
	\begin{minipage}{14pc}
		\includegraphics[width=16pc]{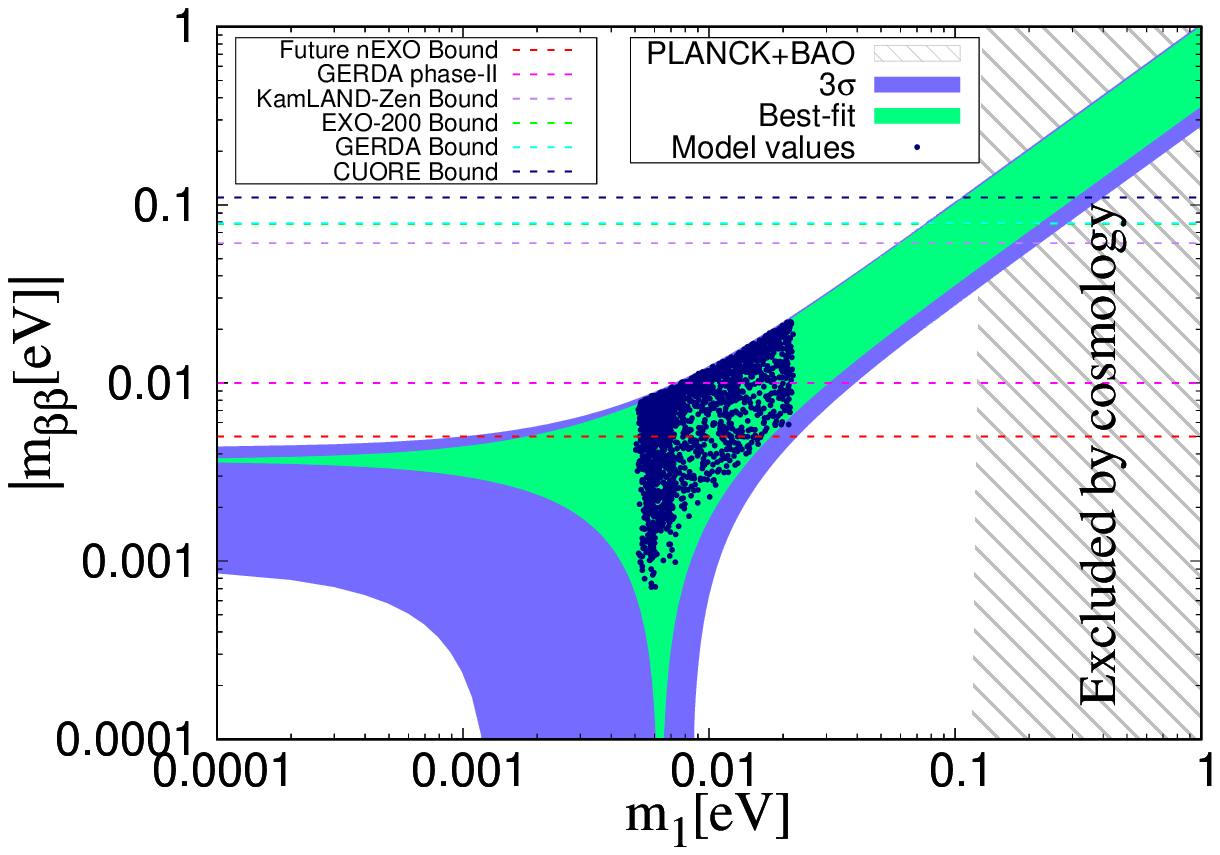}
	\end{minipage}\hspace{2pc}
	\begin{minipage}{14pc}
		\includegraphics[width=16pc]{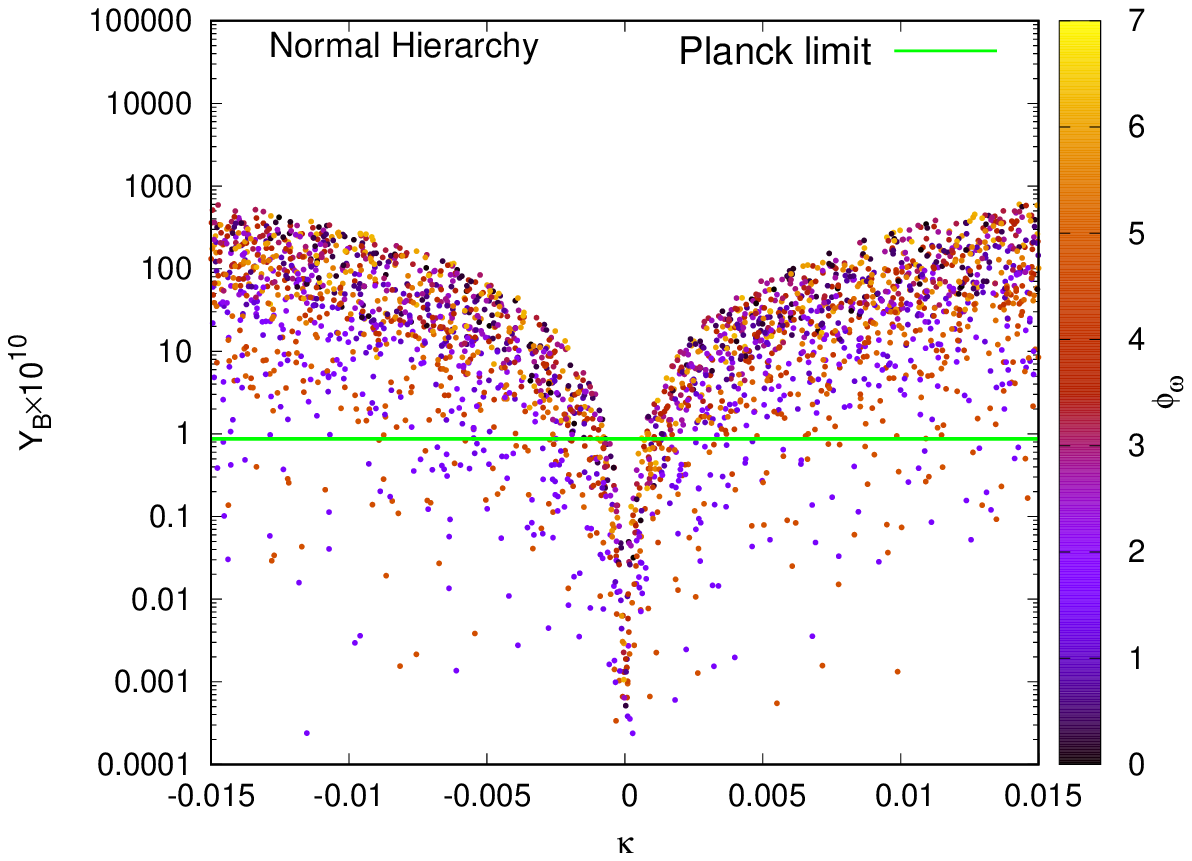}
	\end{minipage}
	\caption{Left panel: $\left\vert m_{\protect\beta \protect\beta }\right\vert $ as a
		function of $m_{1}$. Right panel: $Y_B$ as a function of $\kappa$ while the high energy phase $\phi_{\omega}$ is shown in the colored palette.}
	\label{f1}
\end{figure}
The left panel of Figure (\ref{f1}) shows the correlation between $%
\left\vert m_{\beta \beta }\right\vert $ and the lightest neutrino mass $%
m_{1}$ for the NH. This plot is obtained by varying the
oscillation parameters in their $3\sigma $ range while the Majorana phases $%
\alpha _{21}$ and $\alpha _{31}$\ are varied in the interval $[0\rightarrow
2\pi ]$. The horizontal dashed lines represent the limits on $\left\vert
m_{\beta \beta }\right\vert $ from current $0\nu \beta \beta $ decay
experiments while the vertical gray region is disfavored by the Planck+BAO
data. From this figure, we extract our range of $\left\vert m_{\beta \beta
}\right\vert $%
\begin{equation}
	0.000715\lesssim \left\vert m_{\beta \beta }\right\vert \left( \mathrm{eV}%
	\right) \lesssim 0.022028
	\label{mee}
\end{equation}%
The next-generation
experiments such as GERDA Phase II, CUPID, nEXO and SNO+-II will cover the
values of $\left\vert m_{\beta \beta }\right\vert $ in eq. (\ref{mee}) as
they aim for sensitivities around $ \left( 0.01-0.02\right) \mathrm{eV}$ , $ \left( 0.006-0.017\right) \mathrm{eV}$, $%
 \left( 0.008-0.022\right)
\mathrm{eV}$ and $
\left( 0.02-0.07\right) \mathrm{eV}$, respectively \cite{R36}.

As discussed above, $Y_{B}$
depends on the parameters $\kappa $ and $%
\phi _{\omega }$ which are responsible for the
generation of the observed value of $Y_{B}$, thus we use as their input
ranges $\left[ -0.1\rightarrow 0.1\right] $ and $\left[ 0\rightarrow 2\pi %
\right] $, respectively. Then, we plot in the right panel of figure (\ref{f1}%
) the correlation between $Y_{B}$ and $\kappa $ while the color palette
shows $\phi _{\omega }$. We find that the
observed value of $Y_{B}$ is predicted for the following ranges of $\kappa $
and $\phi _{\omega }$: $\kappa \in \left[ -0.1\rightarrow -0.0085\right] \cup \left[
0.009\rightarrow 0.1\right]$ and $0\lesssim \phi _{\omega }\lesssim
6.279$. We find also that the $CP$ conserving values $\phi _{\omega }=\frac{\pi }{2}$
and $\phi _{\omega }=\frac{3\pi }{2}$ as well as the regions around them are
excluded. Thus, this source of $CP$ violation plays a crucial role in
generating the baryon asymmetry in the present model. A detailed study
concerning the correlation between $Y_{B}$ and the other oscillation parameters has
been performed in ref. \cite{R21}.

\section{Summary and conclusion}
We have studied leptogenesis and $0\nu\beta\beta$ in the framework of a SUSY $SU(5)$ GUT supplemented by a $D_4$ symmetry. We have adopted a neutrino mixing matrix of trimaximal form, and through numerical
analysis we found that the IH scheme is excluded. Interestingly, we found that the predictions for $m_{\beta \beta }$ are accessible in future experiments. On the other hand, by
ignoring flavor effects and taking into account that the three RH neutrinos $%
N_{i}$ are not strongly hierarchical, we computed the baryon asymmetry
parameter $Y_{B}$ in the NH case. This latter depends mainly on the
parameters $\kappa $ and $\phi _{\omega }$ emerged from the correction added
to the Dirac mass matrix. We have shown through plots that
\textrm{generating} the baryon asymmetry in the present model\ calls for non
conserving values of the high energy $CP$ phase $\phi _{\omega }$.

\end{document}